\documentstyle[prl,aps,multicol,epsfig]{revtex}
\begin{document} 
\draft 
\title{Strong coupling theory for the Hubbard model}
\author{A. Dorneich,  M. G. Zacher, C. Gr\"ober and R. Eder}
\address{Institut f\"ur Theoretische Physik, Universit\"at W\"urzburg,
Am Hubland,  97074 W\"urzburg, Germany}
\date{\today}
\maketitle

\begin{abstract}
We reanalyze the Hubbard-I approximation by showing that it is equivalent 
to an effective Hamiltonian describing Fermionic charge fluctuations, which 
can be solved by Bogoliubov transformation. As the most important correction
in the limit of large $U$ and weak spin correlations we augment this
Hamiltonian by further effective particles, which describe composite
objects of a Fermionic charge fluctuation and a spin-, density- or $\eta$ 
excitation. The scheme is valid for positive and negative $U$.
We present results for the single particle Green's function for the 
two-dimensional Hubbard model with and without $t'$ and $t''$ terms, and 
compare to Quantum Monte-Carlo (QMC) results for the paramagnetic phase. The 
overall agreement is significantly improved over the conventional Hubbard-I 
or two-pole approximation.
\end{abstract} 
\begin{multicols}{2}
%%%%%%%%%%%%%%%%%%%%%%%% Intro %%%%%%%%%%%%%%%%%%%%%%%%%%%%%%%%%%%%%%
\section{Introduction}
%%%%%%%%%%%%%%%%%%%%%%%%%%%%%%%%%%%%%%%%%%%%%%%%%%%%%%%%%%%%%%%%%%%%%
The Hubbard model is the simplest system which presumably
incorporates the key features of the strong correlation
limit. Understanding this model will be crucial for
making any progress with cuprate superconductors,
colossal magneto resistance systems or Heavy Fermions.
The special problem in this model is that near half-filling
it represents a dense system of strongly interacting
Fermions, a situation in which a perturbation expansion in $U$
may not be expected to give any meaningful results.
In dealing with this model, one can then pursue two opposing
strategies: one might expect that despite the strong interaction
a perturbation expansion in $U$ remains a meaningful approximation,
and apply conventional many-body theory. The latter means that
one is treating the kinetic energy exactly, and results in the
validity of the Luttinger theorem. On the other hand, by adiabatic continuity
this approach will never produce an insulator at half-filling so that
we can be sure of its breakdown in the limit of large $U/t$.\\
The opposite point of view was taken by Hubbard\cite{hubbard,hubbard1}: 
in his approximations,
the interaction part of $H$ is treated exactly and approximations
are made to the kinetic energy. This results in the breakdown of the
Luttinger theorem, because the physical electron is
effectively split into two particles, one of them corresponding
to an electron moving between empty sites, the other to an
electron moving between sites occupied by an electron of opposite spin. The
energies of formation of these effective particles
differ by $U$, whence for large $U$ the single free-electron band
splits up into two bands formed predominantly
by the two types of effective particles.
Hubbard's approximations, and related schemes like the so-called
$2$-pole approximations\cite{roth,nolting,bernhard,beenen}, 
have been dismissed by some
authors as unphysical, because they do in fact violate the 
Luttinger theorem away from half-filling.
However, in a recent QMC study for the paramagnetic phase of the
Hubbard model\cite{carmarc} we have shown that such criticism is entirely
unwarranted: the Fermi surface, if measured in an `operational way'
from the Fermi energy crossings of the quasiparticle band, indeed
does violate the Luttinger theorem. The doping dependence of the
Fermi surface volume is qualitatively consistent with
Hubbard's results and the main discrepancy being the fact that
one can rather clearly distinguish $4$ `bands' in the spectral functions
rather than the $2$ bands predicted by the Hubbard-I approximation.
More generally, exact diagonalization studies\cite{Dagoreview} tend to produce
for example a photoemission
spectrum consisting of a relatively narrow `quasiparticle band'
(which forms the first ionization states) and an `incoherent continuum'.
A simple two-band form of the spectrum, as produced by the
Hubbard-I approximation, therefore cannot give a quantitative
description of the spectrum.
Motivated by these numerical results we have re-examined the
Hubbard I approximation and attempted to find the most
important corrections to this scheme for the limit of large $U$
and weak spin correlations. We will see that the $4$-band structure
observed in the QMC simulations can be reproduced quite well
by adding two new effective particles which correspond to a composite
object of a `Hubbard quasiparticle' and
a spin-, charge-, or $\eta$-excitation. These
composite objects actually are the quite obvious leading correction
over the Hubbard-I approximation, and we will  see that including them into the
equations of motion leads to an almost quantitative agreement
with the numerical results for a variety of different systems.
%%%%%%%%%%%%%%%%%%%%%%%%%%%%%%%%%%%%%%%%%%%%%%%%%%%%%%%%%%%%%%
\section{A reformulation of the Hubbard I approximation}
%%%%%%%%%%%%%%%%%%%%%%%%%%%%%%%%%%%%%%%%%%%%%%%%%%%%%%%%%%%%%%
We consider the Hubbard Hamiltonian in particle-hole symmetric form,
$H=H_t+H_U$ with:
\begin{eqnarray}
H_t &=& -t \sum_{\langle i,j \rangle} 
(c_{i,\sigma}^\dagger  c_{j,\sigma}^{} + H.c.),
\nonumber \\
H_U &=& U \sum_i (n_{i,\uparrow} - \frac{1}{2})  
(n_{i,\downarrow} - \frac{1}{2}).
\end{eqnarray}
Here ${\langle i,j \rangle}$ denotes summation over all pairs of
nearest neighbors and 
$n_{i,\sigma}^{} = c_{i,\sigma}^\dagger c_{i,\sigma}$.\\
For bipartite lattices one can make use of two
symmetry transformations.
The first one is the particle-hole transformation:
$c_{i,\sigma}^{} \leftrightarrow e^{i \bbox{Q} \cdot \bbox{R}_i}
c_{i,\sigma}^\dagger$, where $\bbox{Q}=(\pi,\pi,\dots,\pi)$.
If the kinetic energy contains only nearest neighbor hopping,
this transformation leaves the Hamiltonian invariant
and exchanges the electron addition spectrum
for momentum $\bbox{k}$ and removal spectrum
for momentum $\bbox{k}+\bbox{Q}$ at half-filling.
Similarly the transformation
$c_{i,\downarrow}^{} \leftrightarrow e^{i \bbox{Q} \cdot \bbox{R}_i}
c_{i,\downarrow}^\dagger$,
$c_{i\uparrow} \rightarrow c_{i\uparrow}$
inverts the sign of $H_U$\cite{micnas}. 
At half-filling this allows to transform solutions
of the positive-U model into those of the negative-U model.
This transformation implies
that the single-particle spectral function at half-filling
is identical for positive and negative $U$.
In the appendix some of the operators introduced in this
work and their transformation properties
under these transformations are listed.\\
We proceed with the calculation of the single-particle 
Green's function. As a first step, following Hubbard\cite{hubbard},
we split the electron annihilation operator into
the two eigenoperators of the interaction part:
\begin{eqnarray}
c_{i,\sigma} &=&
c_{i,\sigma} n_{i,\bar{\sigma}} + 
c_{i,\sigma} (1-n_{i,\bar{\sigma}})
\nonumber \\
&=& \hat{d}_{i,\sigma} + \hat{c}_{i,\sigma}.
\end{eqnarray}
These obey
$ [\hat{d}_{i,\sigma}, H_U] = \frac{U}{2} \hat{d}_{i,\sigma}$
and
$[ \hat{c}_{i,\sigma}, H_U] = - \frac{U}{2} \hat{c}_{i,\sigma}$.
Next, we consider the commutators of these `effective particles' with
the kinetic energy. After some algebra, thereby
using the identity $n_{i,\sigma}$$=$$\frac{n_i}{2} + \sigma S_i^z$,
we find:
\begin{eqnarray}
 [ \hat{c}_{i,\uparrow}, H_t] &=& -t \sum_{j\in N(i)}\; [
 (1-\frac{\langle n\rangle}{2}) c_{j,\uparrow}
+ ( c_{j,\uparrow} S_i^z + c_{j,\downarrow} S_i^-)
\nonumber \\
&&\;\;\;\;\;\; - \frac{1}{2}  c_{j,\uparrow} ( n_i - \langle n\rangle)
+ c_{j,\downarrow}^\dagger c_{i,\downarrow} c_{i,\uparrow} ],
\nonumber \\
\; [ \hat{d}_{i,\uparrow}, H_t] &=& -t \sum_{j\in N(i)}\; [
 \frac{\langle n\rangle}{2} c_{j,\uparrow}
- ( c_{j,\uparrow} S_i^z + c_{j,\downarrow} S_i^-)
\nonumber \\
&&\;\;\;\;\;\; + \frac{1}{2}  c_{j,\uparrow} ( n_i - \langle n\rangle)
- c_{j,\downarrow}^\dagger c_{i,\downarrow} c_{i,\uparrow} ].
\label{commu}
\end{eqnarray}
Here $N(i)$ denotes the $z$ nearest neighbors of site $i$.
Keeping only the first term in each of the square brackets on the
r.h.s. (as we will do for the remainder
of this section) reproduces the Hubbard I approximation.
We specialize to half-filling ($\langle n_{i,\sigma} \rangle = 1/2$) and 
introduce the Green's functions
\begin{equation}
G_{\alpha,\beta}(\vec{k},t)
= -i \langle T \alpha_{\bbox{k},\sigma}^\dagger(t) \beta_{\bbox{k},\sigma}
\rangle,
\label{green}
\end{equation}
where $\alpha,\beta \in \{\hat{c},\hat{d}\}$.
Then, using the anticommutator
relations $\{\hat{d}_{i,\sigma}^\dagger,\hat{d}_{i,\sigma}\}
= n_{i\bar{\sigma}}$,
$\{\hat{c}_{i,\sigma}^\dagger,\hat{c}_{i,\sigma}\}
= (1-n_{i\bar{\sigma}})$, 
$\{ \hat{d}_{i,\sigma}^\dagger,\hat{c}_{i,\sigma}\}=
\{\hat{c}_{i,\sigma}^\dagger,\hat{d}_{i,\sigma}\}=0$
we obtain the following equations of motion:
\begin{eqnarray}
i\partial_t G_{\hat{c},\hat{c}} &=&
\frac{1}{2}\delta(t)
+ \frac{\epsilon_{\bbox{k}}-U}{2}  G_{\hat{c},\hat{c}} + 
\frac{\epsilon_{\bbox{k}}}{2} G_{\hat{d},\hat{c}} , 
\nonumber \\
i\partial_t G_{\hat{d},\hat{c}} &=&
 \frac{\epsilon_{\bbox{k}}}{2} G_{\hat{c},\hat{c}} + 
\frac{\epsilon_{\bbox{k}}+U}{2} G_{\hat{d},\hat{c}},
\nonumber \\
i\partial_t G_{\hat{c},\hat{d}} &=&
\frac{\epsilon_{\bbox{k}}-U}{2} G_{\hat{c},\hat{d}} + 
\frac{\epsilon_{\bbox{k}}}{2} G_{\hat{d},\hat{d}} ,
\nonumber \\
i\partial_t G_{\hat{d},\hat{d}} &=&
\frac{1}{2}\delta(t)+ 
\frac{\epsilon_{\bbox{k}}}{2} G_{\hat{c},\hat{d}} + 
\frac{\epsilon_{\bbox{k}}+U}{2} G_{\hat{d},\hat{d}}  .
\label{simp}
\end{eqnarray}
Taking into account that
the ordinary electron Green's function $G$ and the 
Green's function called $\Gamma$ by Hubbard\cite{hubbard} can be written as
\begin{eqnarray}
G &=& G_{\hat{c},\hat{c}} + G_{\hat{c},\hat{d}} + G_{\hat{d},
\hat{c}} + G_{\hat{d},\hat{d}},
\nonumber \\
\Gamma &=& G_{\hat{d},\hat{c}} + G_{\hat{d},\hat{d}},
\label{hubbI}
\end{eqnarray}
the resulting equations of motions are precisely those
derived in the Hubbard-I approximation:
\begin{eqnarray}
i\partial_t G &=& \delta(t) + (\epsilon_{\bbox{k}}-\frac{U}{2}) G + U \Gamma,
\nonumber \\
i\partial_t \Gamma &=& 
\frac{1}{2}( \delta(t) +\epsilon_{\bbox{k}} G +  U \Gamma ).
\end{eqnarray}
The present formulation, on the other hand,
allows for an appealing physical interpretation
of the Hubbard-I approximation:
we introduce free Fermion operators $h_{\bbox{k},\sigma}^\dagger$ and
$d_{\bbox{k},\sigma}^\dagger$, which correspond to `holes' and `double
occupancies'.
The Hubbard operators are identified with these as follows:
\begin{eqnarray}
\hat{c}_{\bbox{k},\sigma} &\rightarrow& \frac{1}{\sqrt{2}}
h_{-\bbox{k},\sigma}^\dagger
\nonumber \\
\hat{d}_{\bbox{k},\sigma} &\rightarrow& \frac{1}{\sqrt{2}}
d_{\bbox{k},\sigma}^{}.
\label{egon}
\end{eqnarray}
Then, we can formally obtain the set of equations of motion
(\ref{simp}) from the following Hamiltonian
for the holes and double occupancies:
\begin{eqnarray}
H_{eff} &=& \sum_{\bbox{k},\sigma}
\;(\;\frac{\epsilon_{\bbox{k}}+U}{2} d_{\bbox{k},\sigma}^\dagger 
d_{\bbox{k},\sigma}^{}  -
\frac{\epsilon_{\bbox{k}}-U}{2} h_{\bbox{k},\sigma}^\dagger 
h_{\bbox{k}^{},\sigma} \;)
\nonumber \\
&&+  \sum_{\bbox{k}}
(\frac{\epsilon_{\bbox{k}}}{2} \;
d_{\bbox{k},\uparrow}^\dagger h_{-\bbox{k},\downarrow}^\dagger 
+ H.c.) .
\label{hubre}
\end{eqnarray}
The Hamiltonian (\ref{hubre}) is a quadratic form
and readily solved by Bogoliubov transformation:
\begin{eqnarray}
\gamma_{-,\bbox{k},\sigma} &=& u_{\bbox{k}} d_{\bbox{k},\sigma}^{}
+ v_{\bbox{k}} h_{-\bbox{k},\bar{\sigma}}^\dagger,
\nonumber \\
\gamma_{+,\bbox{k},\sigma} &=&-v_{\bbox{k}} d_{\bbox{k},\sigma}^{}
+ u_{\bbox{k}} h_{-\bbox{k},\bar{\sigma}}^\dagger,
\end{eqnarray}
to yield the familiar dispersion relation
\begin{equation}
E_{\pm}(\bbox{k}) = \frac{1}{2}\;[  \epsilon_{\bbox{k}}
\pm \sqrt{ \epsilon_{\bbox{k}}^2 + U^2} \;].
\label{hubdisp}
\end{equation}
To compute the spectral weight of the two bands we use
$c_{\bbox{k},\sigma}= \frac{1}{\sqrt{2}}(
d_{\bbox{k},\sigma}^{} + 
h_{-\bbox{k},\bar{\sigma}}^\dagger)$, whence
\begin{eqnarray}
Z_\pm(\bbox{k}) &=& \frac{1}{2}( u_{\bbox{k}} \mp v_{\bbox{k}} )^2.
\nonumber \\
&=& \frac{1}{2} \left(1 \pm 
\frac{\epsilon_{\bbox{k}}}{\sqrt{\epsilon_{\bbox{k}}^2 + U^2} } \right).
\label{hubweight}
\end{eqnarray}
Again, this is the correct Hubbard-I result.
The above discussion shows the physical content of the Hubbard-I 
approximation: the Hamiltonian (\ref{hubre}) describes Fermionic
particle-like and hole-like charge fluctuations,
created by $d_{\bbox{k},\sigma}^\dagger$
and $h_{-\bbox{k},\bar{\sigma}}^\dagger$, respectively. These
`live' in a background of singly occupied sites.
Particle-like and hole-like charge fluctuations are created in pairs
on nearest neighbors, and individually can hop between nearest neighbors.
The hopping integral for the hole-like particle has opposite
sign as that for the electron-like fluctuation as it has to be, and
the hopping integrals for both
effective particles are $1/2$ times that for the ordinary electrons:
this reflects the fact that due to the
Pauli principle (say) a spin up electron added
to a `background' of singly occupied sites can propagate
to a neighboring site only with a probability
of $1/2$ (we are neglecting any spin correlations of the
background of singly occupied sites).
The factor of $1/\sqrt{2}$ in (\ref{egon}) is due to the fact that
$\langle \hat{c}_{i,\sigma}^\dagger \hat{c}_{i,\sigma}^{}\rangle = 1/2$.
Finally, the particle which stands for the double occupancy has an
energy of formation of $U/2$, the hole-like particle
has an energy of $-U/2$.
As already mentioned,
the Hubbard I approximation therefore describes the splitting
of the physical electron into two new effective particles which
carry with them information
on the `environment' in which the electron has been created.
One of them ($d_{\bbox{k},\sigma}$)
moves between sites occupied by an electron of opposite spin, the other
one ($h_{-\bbox{k},\bar{\sigma}}$)
between empty sites. This is a quite appealing physical idea, but
the above formulation also very clearly highlights the weak points of
the Hubbard-I approximation. In addition to the
mere truncation of the commutators (\ref{commu}), which
is an uncontrolled approximation, these are the following: 
adopting this picture we would have to assume that the states
$h_{i,\uparrow}^\dagger d_{j,\uparrow}^\dagger|0\rangle$
and $h_{i,\downarrow}^\dagger d_{j,\downarrow}^\dagger|0\rangle$
are distinguishable (and in fact orthogonal to one another).
This, however, is in general not the case:
both states have one double occupancy on site $j$ and a hole
on site $i$, and the only difference is that they have been
created in different ways. In fact, using (\ref{egon}) we find
for their overlap
\begin{eqnarray}
\langle 
d_{j,\downarrow} h_{i,\downarrow}
h_{i,\uparrow}^\dagger d_{j,\uparrow}\rangle &=&
-4 \langle S_i^- S_j^+ \rangle
\nonumber \\
&=&- \frac{8}{3} \langle \vec{S}_i \cdot \vec{S}_j \rangle,
\end{eqnarray}
where we have assumed a rotationally invariant ground state in the
second line.
The Hubbard-like approximation scheme thus should work only
for a state with vanishing spin correlations - we will
therefore henceforth assume the spin correlation function
$\langle \vec{S}_i \cdot \vec{S}_j \rangle$ to be zero. \\
A second problem is, that the effective Fermions have to obey a kind of
hard-core constraint - a site cannot be simultaneously occupied
by a hole and a double-occupancy. This constraint is not
accounted for in the derivation of the Hubbard-I approximation:
the equations of motion are obtained from the
Hamiltonian (\ref{hubre})
by treating the $h$ and $d$ as ordinary free Fermion operators.
This problem is the source of the violation of certain sum-rules
when the Hubbard-I approximation is applied in the doped case,
see the discussion given by Avella {\em et al.}\cite{avella}.\\
One last remark is that the commutators
(\ref{commu}) are invariant under the particle-hole transformation,
i.e. they are transformed into each others Hermitian conjugate.
This remains true for the truncated commutators, which give the
Hubbard-I approximation, so that the spectral function
obtained from this is particle-hole symmetric. This can also be verified
directly using (\ref{hubdisp}) and (\ref{hubweight}).
Finally, the spectral function is manifestly invariant under
sign change of $U$, as it has to be.
%%%%%%%%%%%%%%%%%%%%%%%%%%%%%%%%%%%%%%%%%%%%%%%%%%%%%%%%%%%
\section{Extension of the Hubbard I approximation}
\begin{figure}
\epsfxsize=6.0cm
\vspace{-0.0cm}
\hspace{-0.5cm}\epsfig{file=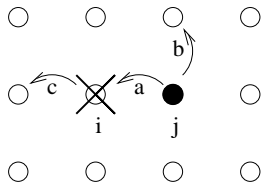,width=6cm,angle=0.0}
\vspace{0.0cm}
\narrowtext
\caption[]{Possible hopping processes which couple
to $\hat{C}_{i,j,\sigma}$. The cross denotes the
spin-, density- or $\eta$-excitation, the black dot the hole.}
\label{fig1} 
\end{figure}

%%%%%%%%%%%%%%%%%%%%%%%%%%%%%%%%%%%%%%%%%%%%%%%%%%%%%%%%%%%
We now want to try and derive an improved version of the 
Hubbard-I approximation. Thereby we will address
neither the problem of nonorthogonality of different
hole/double-occupancy configurations nor the
hard-core constraint - instead, in this work we will restrict ourselves 
entirely to an approximate way of treating the omitted terms
in the commutator relations  (\ref{commu}).
We expect that the present approximation is reasonable
for large $U$ (where the density of holes/double occupancies
is small whence the hard-core constraint is of little
importance) and weak spin correlations (where the nonorthogonality
problem is small). Throughout we stick to the case of
half-filling and no spin polarization, $\langle n_{i,\sigma} \rangle =1/2$.
We return to the basic commutator relations (\ref{commu}) and
consider the terms on the r.h.s. which are omitted in the Hubbard-I
scheme.
The second term in the square bracket is the Clebsch-Gordan
contraction of the spin-1/2 operator $c_{j,\sigma}$ and the
spin-1 operator $\vec{S}_i$ into yet another spin-1/2 operator -
it describes
the coupling of the created hole/annihilated
double occupancy to spin excitations. This term may be expected to
be the most important one in the limit of large positive $U$.
The third term describes in an analogous way the coupling to density
fluctuations, whereas the last term describes the coupling to
the so-called $\eta$-pair excitation\cite{yang}. One may expect that in the
case of {\em negative} $U$ the last two terms are the important ones.\\
In keeping with the basic idea of the Hubbard approximations, namely
to treat the dominant interaction terms exactly, we split also the
composite operator into eigenoperators of the $U$-term and define:
\begin{eqnarray}
\hat{C}_{i,j,\uparrow} &=& 
 \hat{c}_{j,\uparrow} S_i^z + \hat{c}_{j,\downarrow} S_i^-
- \frac{1}{2}\tilde{n}_i\hat{c}_{j,\uparrow}
+ c_{i,\downarrow} c_{i,\uparrow} \hat{d}_{j,\downarrow}^\dagger,
\nonumber \\
\hat{D}_{i,j,\uparrow} &=& 
 \hat{d}_{j,\uparrow} S_i^z + \hat{d}_{j,\downarrow} S_i^-
- \frac{1}{2}\tilde{n}_i\hat{d}_{j,\uparrow}
+ c_{i,\downarrow} c_{i,\uparrow} \hat{c}_{j,\downarrow}^\dagger,
\label{compdef}
\end{eqnarray}
where $\tilde{n}_i= n_i - \langle n \rangle$.
Under the positive/negative $U$ transformation
we have for example $\hat{c}_{j,\uparrow} S_i^z + \hat{c}_{j,\downarrow} S_i^-$
$\rightarrow$ $\frac{1}{2}\tilde{n}_i\hat{d}_{j,\uparrow}
+ c_{i,\downarrow} c_{i,\uparrow} \hat{c}_{j,\downarrow}^\dagger$, i.e.
keeping the at first sight unimportant (for positive $U$)
terms involving density and pairing fluctuations is crucial for
maintaining the exact symmetry under sign change of $U$. We then have
\begin{eqnarray}
[\hat{D}_{i,j,\sigma}, H_U] &=& \frac{U}{2} \hat{D}_{i,j,\sigma},
\nonumber \\
\;[ \hat{C}_{i,j,\sigma}, H_U] &=& - \frac{U}{2} \hat{C}_{i,j,\sigma},
\nonumber \\
\;[ \hat{c}_{i,\uparrow}, H_t] &=& -t \sum_{j\in N(i)}\; [ \;
 \frac{1}{2} c_{j,\uparrow}
+ \hat{C}_{i,j,\sigma} + \hat{D}_{i,j,\sigma} \;],
\nonumber \\
\; [ \hat{d}_{i,\uparrow}, H_t] &=& -t \sum_{j\in N(i)}\; [
 \frac{1}{2} c_{j,\uparrow}
- \hat{C}_{i,j,\sigma} - \hat{D}_{i,j,\sigma} \;].
\label{bigcom}
\end{eqnarray}
The operators $\hat{C}_{i,j,\sigma}$ and
$\hat{D}_{i,j,\sigma}$ may be thought of describing `composite objects'
consisting of a charge fluctuation and a spin-, density- or $\eta$-excitation
on a nearest neighbor.
Ultimately these composite operators carry the quantum numbers of a single
electron, i.e. spin 1/2 and charge $1$.
We also note that under particle-hole transformation
\begin{eqnarray}
\hat{C}_{i,j,\sigma} &\rightarrow& e^{i \bbox{Q}\cdot \bbox{R}_i}
\hat{D}_{i,j,\sigma}^\dagger
\end{eqnarray}
i.e. the composite particles transform in the same way
as the $\hat{c}_{i,\sigma}$ and $\hat{d}_{i,\sigma}$.
Moreover, the commutators (\ref{bigcom})
are invariant under particle-hole transformation,
i.e. this transformation transforms them into each others
Hermitian conjugates.\\
We now enlarge the set of Green's functions
by allowing $\alpha,\beta \in \{\hat{c},\hat{d},
\hat{C},\hat{D}\}$ in (\ref{green}); 
in the language of the `effective Fermions' this means that we are introducing
additional Fermions corresponding to the composite objects.
To obtain a closed system of equations of motion for these
Green's functions,
we need equations for the $G_{\hat{C}\hat{c}}$ and $G_{\hat{D}\hat{c}}$.
As a first step we turn to the commutators $[ \hat{C}_{i,j,\sigma},H_t]$
and  $[ \hat{D}_{i,j,\sigma},H_t]$.
Here we have to distinguish three cases (see Figure \ref{fig1}):\\
a) the hopping term may act along the bond $(i,j)$ and
transport the hole back from $j$
to $i$.\\
b) it may transport the hole even further away from
$i$\\
c) it may transport the spin- density-
or $\eta$-excitation away from site $i$.\\
If we want to restrict ourselves to the $4$ types of operators
$\hat{C}_{i,j,\sigma}$, $D_{i,j,\sigma}$, $\hat{c}_{i,\sigma}$ and
$\hat{d}_{i,\sigma}$, we have to neglect the contributions from the
processes b and c. These processes would produce `strings' of
excitations along a path of length 2 lattice spacings, and we
would have to introduce even more complicated operator
products to describe these.
Restricting ourselves to processes of the the type a,
i.e. replacing $H_t \rightarrow -t \sum_{\sigma}
(c_{i,\sigma}^\dagger c_{j,\sigma} + H.c.)$,
straightforward computation\cite{remark}
gives the  surprisingly simple result
\begin{eqnarray}
[ \hat{C}_{i,j\uparrow} , H_t ]
&=& \frac{3t}{2} \hat{D}_{j,i,\uparrow} + \frac{t}{2} \hat{C}_{j,i,\uparrow} 
-\frac{3t}{4} (\hat{c}_{i,\uparrow} - \hat{d}_{i,\uparrow}),
\nonumber \\
\;[ \hat{D}_{i,j\uparrow} , H_t ]
&=& \frac{3t}{2} \hat{C}_{j,i,\uparrow} + \frac{t}{2} \hat{D}_{j,i,\uparrow} 
-\frac{3t}{4} (\hat{c}_{i,\uparrow} - \hat{d}_{i,\uparrow}).
\label{commu1}
\end{eqnarray}
Again, these relations are particle-hole invariant,
i.e. they are turned into each others Hermitian conjugates by
particle-hole transformation.
In passing we note that
had we reduced the operators $\hat{C}_{i,j\uparrow}$ and
$\hat{D}_{i,j\uparrow}$ to comprise only the terms involving
spin excitations (as might seem appropriate in the case
of large positive $U$), the commutators
would have been much more complicated and in fact the `Hamilton matrix'
$H_k$ to be defined below would have been non-Hermitian.\\
Next, we need the anticommutators
\begin{eqnarray}
 \{ \hat{C}_{i,j,\uparrow}, \hat{c}_{l,\uparrow}^\dagger \}
&=& \delta_{j,l}
\;[ S_i^z S_j^z +S_i^- S_j^+
+ \frac{\tilde{n}_i\tilde{n}_j}{4}
\nonumber \\
&&\;\;\;\;\;\;\;\;\;\;\;\;\;\;\;\;
+ c_{j,\uparrow}^\dagger c_{j,\downarrow}^\dagger
c_{i,\uparrow} c_{i,\downarrow} \rangle \;]
\nonumber \\
&+& \delta_{j,l} \;[
 \frac{1}{2} S_i^z
- \frac{1}{4}\tilde{n}_i
- \frac{1}{2} ( \tilde{n}_i S_j^z +
\tilde{n}_j S_i^z )\;]
\nonumber \\
&+& \delta_{i,l} \;[
\hat{c}_{j,\downarrow} \hat{c}_{i,\downarrow}^\dagger
- \hat{d}_{i,\downarrow} \hat{d}_{j,\downarrow}^\dagger \;]
\nonumber \\
\{ \hat{D}_{i,j,\uparrow}, \hat{c}_{l,\uparrow}^\dagger \}
&=& \delta_{i,l} \;[\;
 \hat{c}_{j,\downarrow} \hat{d}_{i,\downarrow}^\dagger
+\hat{d}_{j,\downarrow}^\dagger \hat{c}_{i,\downarrow}\;]
\end{eqnarray}
Taking the expectation value in the ground state most of the
terms vanish on the basis of symmetries:
$\hat{c}_{j,\downarrow} \hat{d}_{i,\downarrow}^\dagger
+\hat{d}_{j,\downarrow}^\dagger \hat{c}_{i,\downarrow}$ vanishes
due to inversion symmetry of the ground state,
$\hat{c}_{j,\downarrow} \hat{c}_{i,\downarrow}^\dagger
- \hat{d}_{i,\downarrow} \hat{d}_{j,\downarrow}^\dagger$
and $\tilde{n}_i$
vanish due to particle-hole symmetry at half-filling.
All terms containing unpaired spin operators vanish if we assume
that the ground state is 
invariant under spin rotations
(which excludes ferro- or antiferromagnetic solutions).
Finally we obtain:
\begin{eqnarray}
\langle  \{ \hat{C}_{i,j,\uparrow}, 
\hat{c}_{l,\uparrow}^\dagger \} \rangle
&=& \delta_{j,l} \langle
\vec{S}_i \cdot \vec{S}_j + 
\frac{\tilde{n}_i\tilde{n}_j}{4}
+ c_{i,\uparrow}^\dagger c_{i,\downarrow}^\dagger
c_{j,\uparrow} c_{j,\downarrow} \rangle,
\nonumber \\
\langle  \{ \hat{D}_{i,j,\uparrow}, 
\hat{c}_{l,\uparrow}^\dagger \} \rangle
&=& 0.
\label{anticomm}
\end{eqnarray}
It is is easy to see that the expressions
whose expectation values are taken are invariant
under particle-hole transformation, and under the
positive/negative U transformation we have:
\begin{eqnarray}
\frac{\tilde{n}_i\tilde{n}_j}{4}
&\leftrightarrow& S_i^z S_j^z,
\nonumber \\
c_{i,\uparrow}^\dagger c_{i,\downarrow}^\dagger
c_{j,\uparrow} c_{j,\downarrow}
&\leftrightarrow& S_i^+ S_j^-.
\end{eqnarray}
The expectation value of the anticommutator
would be invariant under this positive/negative-U symmetry.\\
For large positive $U$ the terms $\frac{\tilde{n}_i\tilde{n}_j}{4}$
and $_{i,\uparrow}^\dagger c_{i,\downarrow}^\dagger
c_{j,\uparrow} c_{j,\downarrow}$ have a negligible expectation value
and the only important term comes from the spin correlation.
In keeping with our above remarks concerning the role of spin correlations
in the Hubbard I approximation
we will henceforth take the r.h.s. of (\ref{anticomm}) to be zero.
As was discussed above, the Hubbard-I approximation
implicitly assumes $\langle \vec{S}_i \cdot \vec{S}_j \rangle=0$,
and we will therefore keep this value also in (\ref{anticomm}).
We will discuss the consequences of not making this approximation later on.\\
Using the above commutators and (expectation values of) anticommutators
we are now in a position to set up a closed system of equations of motion. 
In the following we give explicit formulas only for a 1D chain
with nearest-neighbor hopping, but the generalization to higher
dimensions and/or longer range hopping integrals will be self-evident.
We introduce the Fourier transforms
\[
\hat{C}_{\pm,\sigma}(\bbox{k}) =
\sqrt{\frac{4}{3N}} \sum_j e^{i \vec{k}\cdot\vec{R}_j}
\hat{C}_{j,j\pm 1,\sigma},
\]
(and analogously for the $\hat{D}$'s) and
define the vector
$\vec{G}_{c}=(G_{\hat{c}\hat{c}},G_{\hat{d}\hat{c}},
G_{\hat{C}+,\hat{c}},G_{\hat{C}-,\hat{c}},
G_{\hat{D}+,\hat{c}},G_{\hat{D}-,\hat{c}})$.
Here $\hat{C}\pm$ is shorthand for $\hat{C}_{\pm,\sigma}(\bbox{k})$.
Combining (\ref{commu}), (\ref{compdef}) and (\ref{commu1}),
and performing a spatial Fourier transformation
the equations of motion are readily found to be
\begin{equation}
(i\partial_t - H_k) \vec{G}_c = \delta(t) B_c
\label{eqnsys}
\end{equation}
where the Hermitian  matrix $H_k$ is given by
\end{multicols}
\[
H_k = \left(
\begin{array}{c c c c c c c c c c c}
\frac{\epsilon_{\bbox{k}}-U}{2}&,& \frac{\epsilon_{\bbox{k}}}{2}&,
& -\tilde{t} e^{-ik_x/2}&,& -\tilde{t} e^{ik_x/2}&,
& -\tilde{t} e^{-ik_x/2}&,& -\tilde{t} e^{ik_x/2}\\
\frac{\epsilon_{\bbox{k}}}{2}&,& \frac{\epsilon_{\bbox{k}+U}}{2}&,
&  \tilde{t} e^{-ik_x/2}&,&  \tilde{t} e^{ik_x/2}&,
&  \tilde{t} e^{-ik_x/2}&,&  \tilde{t} e^{ik_x/2}\\
-\tilde{t} e^{ik_x/2}&,&  \tilde{t} e^{ik_x/2}&,&
-\frac{U}{2}&, & \frac{t}{2}&, & 0&, & \frac{3t}{2} \\
-\tilde{t} e^{-ik_x/2}&,&  \tilde{t} e^{-ik_x/2}&,&
\frac{t}{2}&, & -\frac{U}{2}&, & \frac{3t}{2}&, & 0 \\
-\tilde{t} e^{ik_x/2}&,&  \tilde{t} e^{ik_x/2}&,&
 0&, & \frac{3t}{2}&,& \frac{U}{2}&, & \frac{t}{2} \\
-\tilde{t} e^{-ik_x/2}&,&  \tilde{t} e^{-ik_x/2}&,&
 \frac{3t}{2}&, & 0&, & \frac{t}{2}&, & -\frac{U}{2} \\
\end{array}
\right)
\]
\begin{multicols}{2}
\noindent
and $B_c=(\frac{1}{2},0,0,0,0,0)$.
The equation system (\ref{eqnsys} ) can be solved for each momentum and
frequency by using the spectral resolution of the Hamilton matrix $H_k$.
This yields
the Green's functions $G_{\hat{c}\hat{c}}$ 
and $G_{\hat{d}\hat{c}}$ for each momentum and frequency.
In an analogous way we can also derive an equation system
for $G_{\hat{c}\hat{d}}$ and $G_{\hat{d}\hat{d}}$. Thereby the 
matrix $H_k$ stays unchanged,
whereas the r.h.s. is changed into $B_d=(0,1/2,0,0,0,0)$.
Finally, the full electron Green's function is obtained by adding
up the four Green's functions according to (\ref{hubbI}).
Upon forming the Laplace transform $G(\bbox{k},z)$
we finally obtain the spectral density 
\begin{equation}
A(\bbox{k},\omega) = -\lim_{\eta\rightarrow 0} \frac{1}{\pi}
\Im G(\bbox{k},\omega + i \eta).
\end{equation}
In passing we note that this way of computing  $A(\bbox{k},\omega)$
guarantees the validity of the sum rule
\begin{equation}
\int_{-\infty}^{\infty} A(\bbox{k},\omega) d\omega = 1.
\label{sumrule1}
\end{equation}
Since the particle-hole symmetry of the relations (\ref{bigcom}),
(\ref{commu1}) and (\ref{anticomm}) in turn
guarantees particle-hole symmetry of the entire
spectral function, we find that the sum-rule for the
particle number is fulfilled automatically:
\begin{equation}
\sum_{\bbox{k}} \int_{-\infty}^0  A(\bbox{k},\omega) d\omega
= \frac{N}{2}.
\label{sumrule2}
\end{equation}
As a last remark we note that
going over to higher dimensions
or adding longer ranged hopping integrals
poses no problem whatsoever -
for each spatial dimension $\alpha$
we have to add four additional rows and columns
containing the $\hat{C}_{i,j}$ and $\hat{D}_{i,j}$ 
where $i$ and $j$
are nearest neighbors in $\pm \alpha$-direction.
Similarly, if we add an additional hopping integral $t'$
between $2^{nd}$ or $3^{rd}$ nearest neighbors (the number of whom we denote
by $z'$) to the Hamiltonian, we have to
add $2z'$ rows and columns, containing the
$\hat{C}_{i,j}$ and $\hat{D}_{i,j}$ 
with $2^{nd}$ or $3^{rd}$ nearest neighbors $i$ and $j$.
In each case these additional rows and columns contain only mixing 
terms amongst themselves or with
$G_{\hat{c},\hat{c}}$ 
and $G_{\hat{d}\hat{c}}$, so that the extension is completely trivial.
%%%%%%%%%%%%%%%%%%%%%%%%%%%%%%%%%%%%%%%%%%%%%%%%%
\section{Comparison with numerics}
\begin{figure}
\epsfxsize=6.0cm
\vspace{-0.0cm}
\hspace{-0.5cm}\epsfig{file=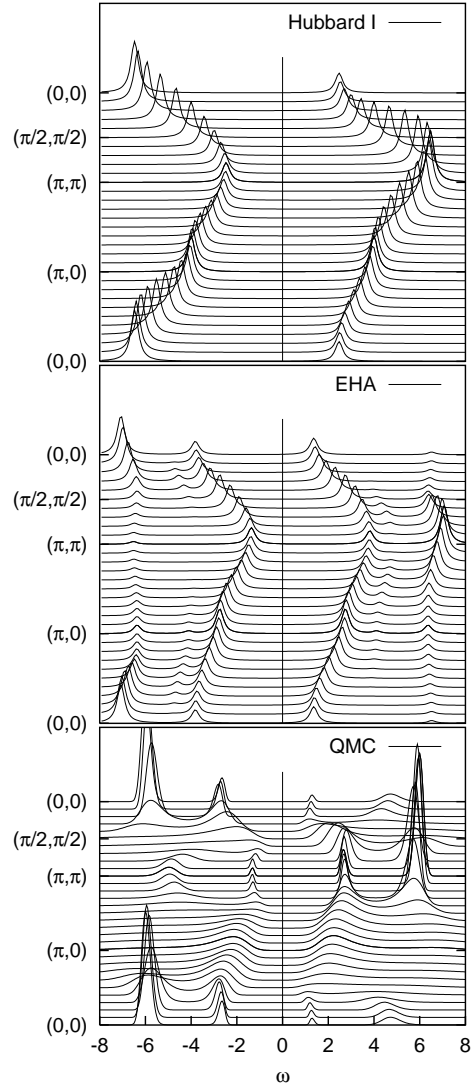,height=6.5cm,angle=270.0}
\vspace{0.5cm}
\narrowtext
\caption[]{Single particle spectral function for the
Hubbard Model with $U/t=8$ from the Hubbard I approximation,
the extended Hubbard approximation, and QMC simulations on a $20\times 20$
lattice at temperature $T=t/3$. In this as well as in all
following figure, the approximate spectra have been given an 
artificial Lorentzian broadening $\eta=0.20t$.
To compensate for the stronger broadening
the QMC spectra have been multiplied by an additional factor of $2$.}
\label{fig2} 
\end{figure}

\begin{figure}
\epsfxsize=6.0cm
\vspace{-0.0cm}
\hspace{-0.5cm}\epsfig{file=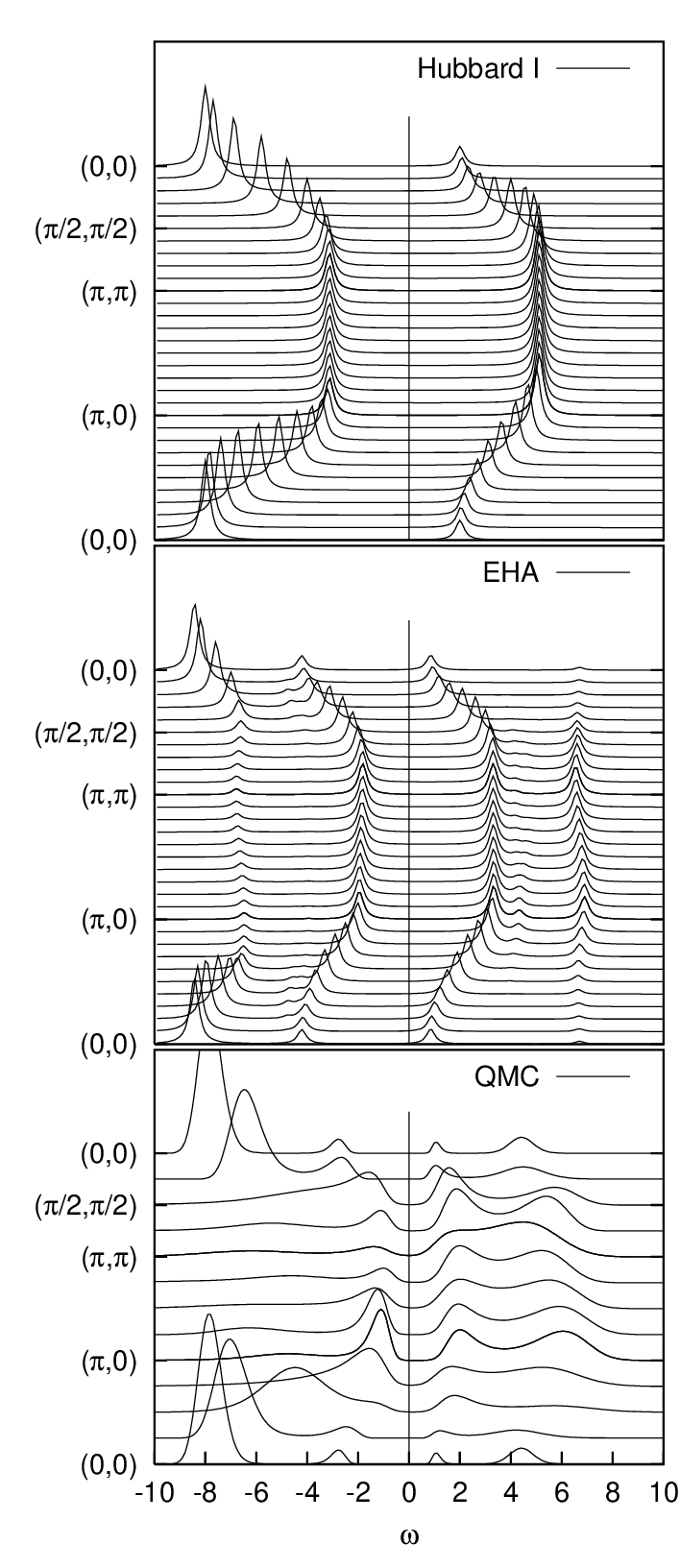,height=6.5cm,angle=270.0}
\vspace{0.5cm}
\narrowtext
\caption[]{Single particle spectral function for the
Hubbard Model with $U/t=8$, $t'=t/2$ from the Hubbard I approximation,
the extended Hubbard approximation, and QMC simulations on an $8\times 8$
lattice at temperature $T=t/3$. To compensate for the stronger broadening
the QMC spectra have been multiplied by an additional factor of $4$.}
\label{fig3} 
\end{figure}

\begin{figure}
\epsfxsize=6.0cm
\vspace{-0.0cm}
\hspace{-0.5cm}\epsfig{file=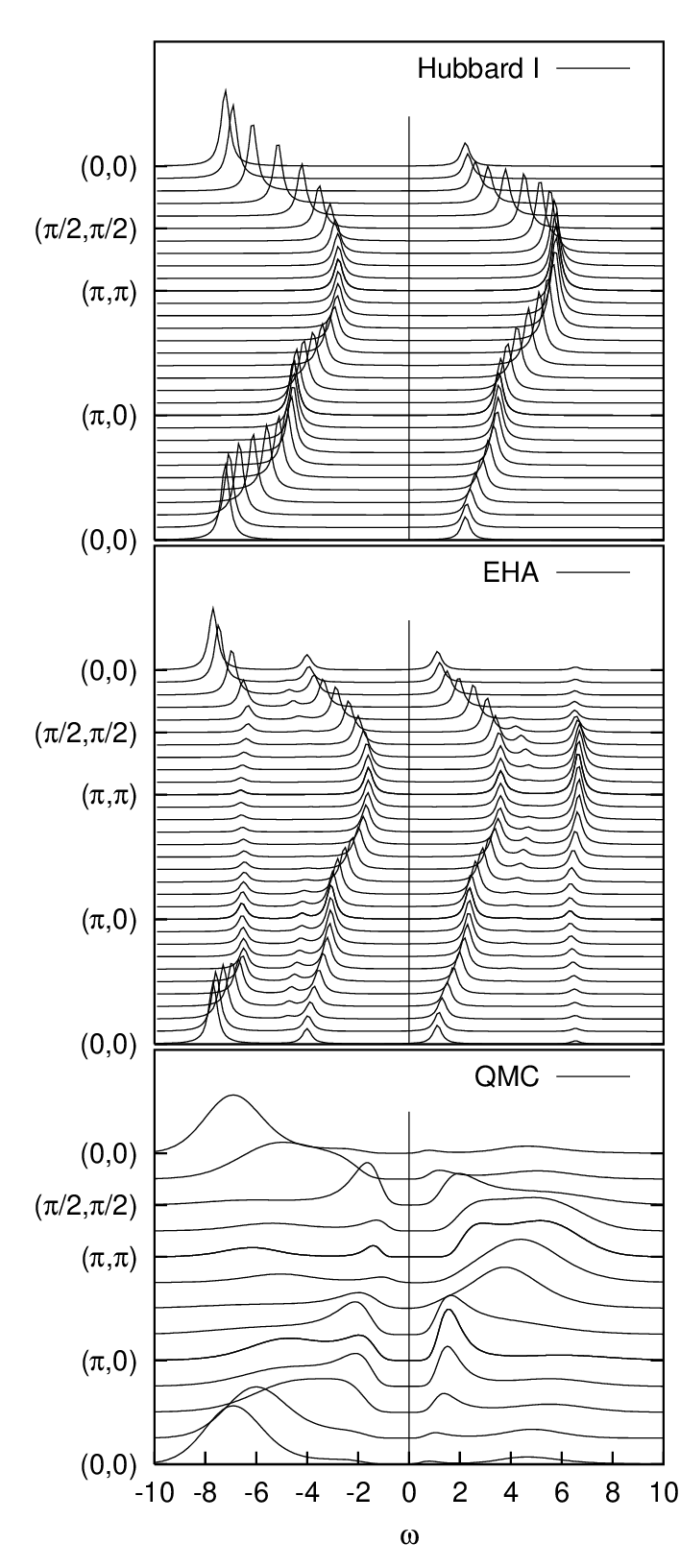,height=6.5cm,angle=270.0}
\vspace{0.5cm}
\narrowtext
\caption[]{Single particle spectral function for the
Hubbard Model with $U/t=8$, $t''=t/4$ from the Hubbard I approximation,
the extended Hubbard approximation, and QMC simulations on an $8\times 8$
lattice at temperature $T=t/3$. To compensate for the stronger broadening
the QMC spectra have been multiplied by an additional factor of $4$.}
\label{fig4} 
\end{figure}

\begin{figure}
\epsfxsize=6.0cm
\vspace{-0.0cm}
\hspace{-0.5cm}\epsfig{file=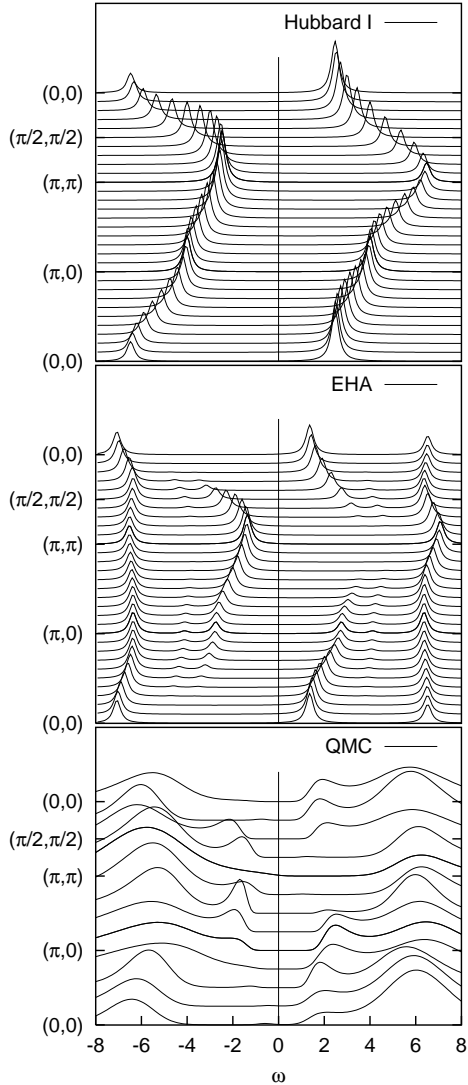,height=6.5cm,angle=270.0}
\vspace{0.5cm}
\narrowtext
\caption[]{Spectral function of the $\tilde{c}$-operator for the
Hubbard Model with $U/t=8$ from the Hubbard I approximation,
the extended Hubbard approximation, and QMC simulations on an $8\times 8$
lattice at temperature $T=t/3$. To compensate for the stronger broadening
the QMC spectra have been multiplied by an additional factor of $4$.}
\label{fig5} 
\end{figure}

\begin{figure}
\epsfxsize=6.0cm
\vspace{0.0cm}
\hspace{-0.5cm}\epsfig{file=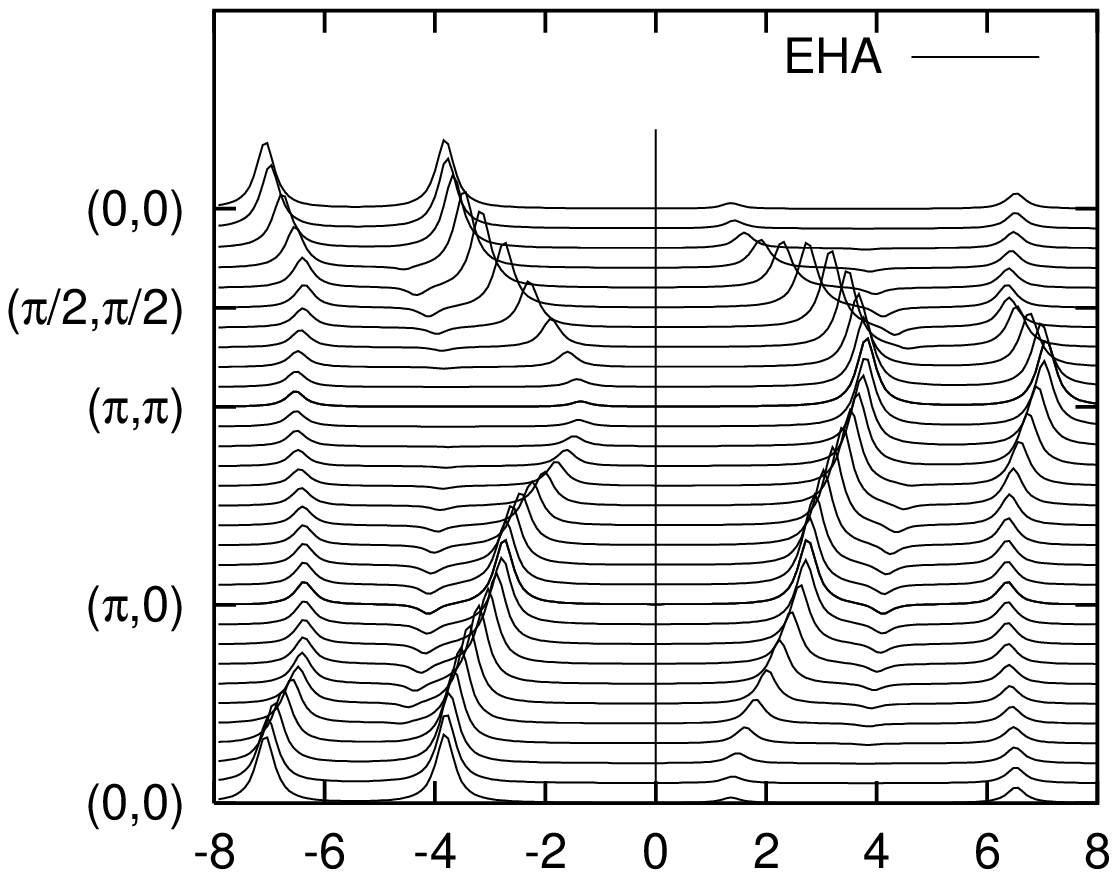,height=6.5cm,angle=270.0}
\vspace{0.5cm}
\narrowtext
\caption[]{Spectral function calculated with the extended
Hubbard approximation for $x=-0.2$.}
\label{fig6} 
\end{figure}

%%%%%%%%%%%%%%%%%%%%%%%%%%%%%%%%%%%%%%%%%%%%%%%%%
Following the discussion in the preceding
section we can calculate the full electron Green's function,
including the (presumably) most important corrections over the
Hubbard I approximation in the limit of
weak spin correlations and large $U$. We now proceed to a comparison of the
obtained results results for the
spectral density $A(\bbox{k},\omega)$,
with the spectral density obtained
from Quantum Monte-Carlo (QMC) simulations.
Thereby the temperature for the QMC simulation, $T=0.33t$,
was chosen such that the spin correlation length
is only approximately $1.5$ lattice spacings - the results thus
are probably quite representative for the paramagnetic
regime which our approximation aims to describe.
Moreover, the value of $U/t=8$ is already rather large,
so that we may also hope to have a small density of
holes/double occupancies and the neglect of the hard-core constraint
be justified. As a general remark concerning the QMC spectra
we note that the MaxEnt procedure used for the analytic continuation
to the real axis es most reliable for `features' with large
weight - this means that the position of tiny peaks is less accurate than that
of large ones.\\
Figure \ref{fig2} then compares the spectral density
obtained from the  Hubbard I approximation, our
extended Hubbard approximation (EHA) and QMC simulation.
The Hubbard I approximation gives only a relatively crude
fit to the actual spectral density obtained by QMC.
The extended Hubbard approximation, on the other hand,
gives an all in all
quite correct description of the spectral density.
Out of the  $10=2+4+4$ bands produced by diagonalizing
$H_k$, only four bands do have an appreciable spectral weight.
These four intense bands correspond rather well to 
$4$ broad `bands' of intense spectral weight which
can be roughly identified in the QMC result. 
It is interesting to note that a recent strong-coupling expansion for the
Hubbard model by Pairault {\em et al.}\cite{Pairault}
also produced a $4$-band structure, although only for the 1D model.
The dispersion of the spectral weight along the bands is
also reproduced quite satisfactorily be the extended Hubbard approximation.
The main difference is the apparently strongly
$\bbox{k}$-dependent width of the spectra
produced by QMC, which however is outside the
scope of the present approximation, which (in the limit
$\eta \rightarrow 0$) produces sharp $\delta$-peaks
without any broadening. A more severe deficiency
of the EHA is, that it tends to predict
too high excitation energies, resulting in a somewhat
too large value of the
Hubbard gap. In any way, however, the magnitude of the Hubbard gap comes out 
better than in the Hubbard-I approximation.\\
We proceed to the Hubbard model with an additional
hopping integral $t'$ between $2^{nd}$ nearest (i.e. (1,1)-like) neighbors.
Figure \ref{fig3} again compares the Hubbard I approximation,
the EHA, and the results of a QMC simulation on an $8\times 8$ lattice.
The agreement between EHA and the QMC result is again quite good, the main
discrepancy being again an overall overestimation of
the binding energies. On the other hand, the apparent
$4$-band structure, the dispersion of the
peak energies and the spectral weight agrees well with the 
numerical result. In particular, the
$2$ bands in inverse photoemission (i.e. $\omega>0$)
predicted by the EHA can be seen very clearly in the QMC spectra).
All in all the agreement is even better than for the case $t'=0$, which most 
probably is due to the fact
that the spin correlations are weaker with $t'$, so that the
assumption of an uncorrelated spin state is better justified
in this case.\\
We proceed to the case of a hopping integral integral $t''$ between 
$3^{rd}$ nearest (i.e. (2,0)-like) neighbors.
Here we have chosen $t''/t=0.25$, because for the larger value
$t''/t=0.5$ the QMC simulation still predicted a metallic state
at $U/t=8$. Figure \ref{fig4} shows the spectral functions for $t''/t=0.25$.
Again, we can roughly identify $4$ bands
and there is reasonable agreement for the
dispersion. The weight of the quasiparticle band in photoemission
near $(\pi,\pi)$ (at $\omega \approx -2$) is not reproduced
very well by the EHA, but again the ubiquitous $4$-band structure
is rather clearly visible.\\
Next, we turn to a somewhat indirect check of the
approximation.
The ordinary electron creation operator is the symmetric
combination of the Hubbard operators.
However, we might also define the antisymmetric
combination
\begin{equation}
\tilde{c}_{i,\sigma} = c_{i,\sigma} n_{i,\bar{\sigma}}
- c_{i,\sigma} (1-n_{i,\bar{\sigma}}).
\end{equation}
This operator has the same quantum numbers as the
original electron operator, and therefore obeys the same
selection rules. It follows that when acting on the
ground state, this operator probes the same final states
as the electron operator, the only difference being the
matrix element viz. the spectral weight of the respective peak
in the spectral density.
In fact the  Green's function
\begin{equation}
\tilde{G}(\bbox{k},t)
= -i \langle T \tilde{c}_{\bbox{k},\sigma}^\dagger(t) 
\tilde{c}_{\bbox{k},\sigma}
\rangle
\end{equation}
can be expressed as
\begin{equation}
\tilde{G} = G_{cc} - G_{dc} - G_{cd} + G_{dd}.
\end{equation}
It therefore is easy to calculate within our approximation, and comparison
with the QMC result provides an additional check for our
description of the electronic structure. Note that 
the operator $\tilde{c}_{\bbox{k},\sigma}$ enhances the practically
dispersionless bands at $\omega=\pm 6$ - since these
bands now have a larger weight, their position and dispersion
are more reliable than in the `ordinary' photoemission spectrum.
This is shown in Figure \ref{fig5}, where indeed quite good
agreement is found between the EHA and the numerical result.\\
Finally, we turn to the discussion of choosing a nonvanishing
expectation value of the anticommutator in (\ref{anticomm}),
i.e. we assume that $\langle \{ \hat{C}_{i,j,\sigma}, 
\hat{c}_l^\dagger\} \rangle = \delta_{j,l} x \neq 0$.
This changes the rhs of the equation system (\ref{eqnsys})
to $B_c = (1/2, 0, \sqrt{\frac{4}{3}}x, \dots)$ but leaves the matrix
$H_k$ unchanged. In other words, the {\em dispersion} of the
bands, which is determined by the eigenvalues of $H_k$ stays unchanged
and only the spectral weight of the peaks changes.
Moreover the sum rules (\ref{sumrule1}) and
(\ref{sumrule2}) retain their validity also in this case.
For large positive $U$ and $n=1$ charge fluctuations will be strongly
suppressed and double occupancies will have a small probability,
so that the dominant contribution to $x$ comes from the spin correlation
function $\langle \vec{S}_i \cdot \vec{S}_j \rangle $, whence
we should choose $x<0$.  Assuming for example
a negative $x$ of moderate value,
$x=-0.2$, then leads to little change in the
calculated spectral density (see Figure \ref{fig6}): 
the same two bands which had a large
spectral weight for $x=0$ retain a large spectral weight also
in this case. There is, however, a rather undesirable feature
associated with the bands with small spectral weight:
numerical evaluation shows, that for some regions of $\bbox{k}$-space
these bands acquire a small but {\em negative} weight.\\
The physical origin of this problem is probably related to nonorthogonalities
of basis states: in principle we could interpret the
matrix $H_k$ as a Hamiltonian describing 
(in 2D with only nearest neighbor hopping)
$6$ types of Fermionic `effective particles'.
Quite generally, the anticommutator-relation
$\{ a, b^\dagger \} = x \neq 0$ implies that the
wave functions corresponding to the
Fermi operators $a^\dagger$ and $b^\dagger$ are non-orthogonal.
While an exact overlap matrix can never have
negative eigenvalues but at most develop zero eigenvalues
(indicating that the set of basis states is overcomplete),
any approximation to the matrix elements may
lead, as an artefact of the approximation, to
negative eigenvalues. 
Since we are using only approximate values for the
overlaps, it may happen that we obtain states
with a nominally negative norm, whence we can
get poles of negative weight.
Setting $x=0$ throughout amounts to assuming that all
our effective particles are orthogonal to one another and
obviously removes the problem with nonorthogonalities.
This seems reasonable, because we are neglecting overlap terms
proportional to the spin correlation function already in the
Hubbard-I approximation.
The lesson then is basically the same as discussed already
before: the Hubbard-I approximation is well defined
only when applied to an `ideally paramagnetic' state with no correlations
of finite range, and applying it to a state with
finite spin correlations represents an approximation.
%%%%%%%%%%%%%%%%%%%%%%%%%%%%%%%%%%%%%%%%%%%%%%%%%%%%%%%%%%%
\section{Conclusion}
%%%%%%%%%%%%%%%%%%%%%%%%%%%%%%%%%%%%%%%%%%%%%%%%%%%%%%%%%%%
In summary, we have investigated the most important
corrections over the Hubbard-I approximation in the limit
$U/t\rightarrow \infty$ and electron density $n=1$.
We have seen that the Hubbard-I approximation
describes charge fluctuations on a `background' of
singly occupied sites, which is moreover assumed to have zero spin 
correlations. The charge fluctuations are point-like, and
correspond to an electron moving between empty sites and
an electron moving between singly occupied sites.
We note that a very similar construction can also be applied to the Kondo
lattice\cite{oana} and in fact 
reproduces the single particle spectra very well.
This is probably due to the fact that the
Kondo lattice has a unique ground state in the limit of zero
kinetic energy, whereas the ground state of the Hubbard model
is highly degenerate in the case $t=0$.\\
In our extended scheme for the Hubbard model we have augmented these 
point-like charge fluctuations by additional `particles'
which are composite in character and consist of
a point-like charge fluctuation together with a spin-, density-,
or $\eta$-excitation. 
Comparison of the obtained single-particle spectral density
with QMC results for a variety of systems
showed a quite reasonable agreement. In particular
the apparent $4$-band structure seen in the numerical spectra finds
its natural explanation in the extended Hubbard approximation.
We also note that QMC simulations where the spectra of the composite
excitations have actually been computed\cite{carsten}, further support
the present interpretation. We thus have a quite successful method
of computing the full quasiparticle band structure of the
Hubbard model, at least in the paramagnetic case and at half-filling.\\
The present scenario for the nature of the composite
excitations also allows to make a connection with various theories
for the hole motion in an 
antiferromagnet\cite{Bulaevski,Trugman,Eder,Sushkov,Reiter,Hayn}. 
There, one is
describing holes dressed by antiferromagnetic spin fluctuations.
When acting on the N\'eel state
the operators $\hat{C}_{i,j,\sigma}$ obviously describe precisely a
hole together with a `spin wave' on a nearest neighbor or,
put another way, a `string' of length one.
The terms which were omitted from the equation of motion for the
$\hat{C}_{i,j,\sigma}$ then would correspond to strings of length two
and so on. While such longer-ranged strings
are apparently of minor importance in the paramagnetic phase,
one may expect that they become more and more important for
the description of the dispersion the stronger the
antiferromagnetic correlations. The relative
importance of such longer ranged strings
therefore may be the mechanism for the
crossover from the Hubbard-I like dispersion in the
paramagnetic phase to the spin-density-wave like dispersion
in the antiferromagnetic phase. Similarly, one might think of formulating
the entire Hubbard-I approximation also in the antiferromagnetic
phase, by constructing the Hamiltonian for charge
fluctuations explicitly for a N\'eel ordered
spin background.\\
We thank H.-G. Evertz and W. Hanke for discussions.
This work was supported by BMBF (05SB8WWA1),
computations were performed at HLRS Stuttgart, LRZ M\"uchen and HLRZ J\"ulich.
\section{Appendix}
Transformation properties of various operators under
particle-hole and positive/negative $U$ transformation.\\[0.5cm]
\begin{tabular}{|c| c| c|}
\hline
Operator & particle-hole & positive/negative $U$\\
\hline
$\hat{c}_{i,\uparrow} $&$ 
e^{i \bbox{Q} \cdot \bbox{R}_i } \hat{d}_{i,\uparrow}^\dagger $&$
\hat{d}_{i,\uparrow} $\\
$\hat{c}_{i,\downarrow} $&$ 
e^{i \bbox{Q} \cdot \bbox{R}_i } \hat{d}_{i,\downarrow}^\dagger $&$
e^{i \bbox{Q} \cdot \bbox{R}_i } \hat{c}_{i,\downarrow}^\dagger $\\
$\hat{d}_{i,\uparrow} $&$ 
e^{i \bbox{Q} \cdot \bbox{R}_i } \hat{c}_{i,\uparrow}^\dagger $&$
\hat{c}_{i,\uparrow} $\\
$\hat{d}_{i,\downarrow} $&$ 
e^{i \bbox{Q} \cdot \bbox{R}_i } \hat{c}_{i,\downarrow}^\dagger $&$
e^{i \bbox{Q} \cdot \bbox{R}_i } \hat{d}_{i,\downarrow}^\dagger $\\
$S_i^+ $&$ - S_i^- $&$ e^{i \bbox{Q} \cdot \bbox{R}_i } c_{i,\uparrow}^\dagger
 c_{i,\downarrow}^\dagger $\\
$S_i^- $&$ - S_i^+ $&$ e^{i \bbox{Q} \cdot \bbox{R}_i }
  c_{i,\downarrow}  c_{i,\uparrow} $\\
$S_i^z $&$ -S_i^z $&$ \frac{1}{2}(n_i - 1) $\\
$n_i - 1 $&$ 1 - n_i $&$ 2 S_i^z $\\
$\hat{C}_{i,j,\uparrow} $&$ e^{i \bbox{Q} \cdot \bbox{R}_i }
\hat{C}_{i,j,\uparrow}^\dagger $&$ -\hat{D}_{i,j,\uparrow} $\\
$\hat{C}_{i,j,\downarrow} $&$ e^{i \bbox{Q} \cdot \bbox{R}_i }
\hat{C}_{i,j,\downarrow}^\dagger $&$ -e^{i \bbox{Q} \cdot \bbox{R}_i }
\hat{D}_{i,j,\uparrow}^\dagger $\\
\hline
\end{tabular}
 
%\end{multicols}
%%%%%%%%%%%%%%%%%%%%%%%%%%%%%%%%%%%%%%%%%%%%%%%%%%%%%%%%%%%%
%%%%%%%%%%%%%%%%%%%%%%%%%%%%%%%%%%%%%%%%%%%%%%%%%%%%%%%%%%%%
%%%%%%%%%%%%%%%%%%%%%%%%%%%%%%%%%%%%%%%%%%%%%%%%%%%%%%%%%%%%
%%%%%%%%%%%%%%%%%%%%%%%%%%%%%%%%%%%%%%%%%%%%%%%%%%%%%%%%%%%%
%%%%%%%%%%%%%%%%%%%%%%%%%%%%%%%%%%%%%%%%%%%%%%%%%%%%%%%%%%%%
%%%%%%%%%%%%%%%%%%%%%%%%%%%%%%%%%%%%%%%%%%%%%%%%%%%%%%%%%%%%
%%%%%%%%%%%%%%%%%%%%%%%%%%%%%%%%%%%%%%%%%%%%%%%%%%%%%%%%%%%%
%%%%%%%%%%%%%%%%%%%%%%%%%%%%%%%%%%%%%%%%%%%%%%%%%%%%%%%%%%%%
\vspace{10cm}
%%%%%%%%%%%%%%%%%%%%%%%%%%%%%%%%%%%%%%%%%%%%%%%%%%%%%%%%%%%%
%%%%%%%%%%%%%%%%%%%%%%%%%%%%%%%%%%%%%%%%%%%%%%%%%%%%%%%%%%%%
%%%%%%%%%%%%%%%%%%%%%%%%%%%%%%%%%%%%%%%%%%%%%%%%%%%%%%%%%%%%
\end{multicols}

\begin{references}
\bibitem{remark}
 The more complicated (anti)commutators in this work were obtained by
 a Mathematica implementation of the Fermion algebra.
\end{references}
\end{document}